\documentclass[journal]{IEEEtran}
\usepackage{amssymb}
\usepackage{amsmath}
\usepackage{amssymb}
\usepackage{epsfig}
\usepackage{multirow}
\usepackage{url}
\usepackage{bm}
\usepackage{verbatim}
\usepackage[linkcolor=black,anchorcolor=blue,citecolor=black]{hyperref}
\usepackage{graphics,setspace}
\usepackage{graphicx}
\usepackage[caption=false,font=small]{subfig}
\usepackage{CJK}
\usepackage[numbers,sort&compress]{natbib}

\makeatletter
\newcommand{\thickhline}{%
	\noalign {\ifnum 0=`}\fi \hrule height 1pt
	\futurelet \reserved@a \@xhline
}

\usepackage[dvipsnames]{xcolor}

\definecolor{mypink1}{rgb}{0.858, 0.188, 0.478}
\definecolor{mypink2}{RGB}{219, 48, 122}
\definecolor{mypink3}{cmyk}{0, 0.7808, 0.4429, 0.1412}
\definecolor{mygray}{gray}{0.6}

\ifCLASSINFOpdf
\else
\fi

\begin{document}
%
\title{Any-to-Many Voice Conversion with Location-Relative Sequence-to-Sequence Modeling}
%
%

\author{Songxiang~Liu,~\IEEEmembership{Student~Member,~IEEE,}\thanks{This project is partially supported by a grant from the HKSAR Government's Research Grants Council General Research Fund (Project no. 14208817).} \thanks{Songxiang Liu, Yuewen Cao, Disong Wang, Xunying Liu and Helen Meng are with the Human-Computer Communications Laboratory (HCCL), the Department of Systems Engineering and Engineering Management, The Chinese University of Hong Kong, Hong Kong SAR, China.} \thanks{Xixin Wu is with Engineering Department, Cambridge University, UK.}
        Yuewen Cao,~\IEEEmembership{Student~Member,~IEEE,}
        Disong Wang,~\IEEEmembership{Student~Member,~IEEE,}
        Xixin Wu$^*$,~\IEEEmembership{Member,~IEEE,}\thanks{* Corresponding author, contact e-mail: xw369@cam.ac.uk}
        Xunying Liu,~\IEEEmembership{Member,~IEEE,}
        and~Helen~Meng,~\IEEEmembership{Fellow,~IEEE} }
        

%
%

\markboth{IEEE/ACM TRANSACTIONS ON AUDIO, SPEECH, AND LANGUAGE PROCESSING, VOL. 29, 2021}%
{Shell \MakeLowercase{\textit{et al.}}: Any-to-Many Voice Conversion with Location-Relative Sequence-to-Sequence Modeling}
%



\maketitle

\begin{abstract}
This paper proposes an any-to-many location-relative, sequence-to-sequence (seq2seq), non-parallel voice conversion approach, which utilizes text supervision during training.
In this approach, we combine a bottle-neck feature extractor (BNE) with a seq2seq synthesis module. During the training stage, an encoder-decoder-based hybrid connectionist-temporal-classification-attention (CTC-attention) phoneme recognizer is trained, whose encoder has a bottle-neck layer. A BNE is obtained from the phoneme recognizer and is utilized to extract speaker-independent, dense and rich spoken content representations from spectral features. Then a multi-speaker location-relative attention based seq2seq synthesis model is trained to reconstruct spectral features from the bottle-neck features, conditioning on speaker representations for speaker identity control in the generated speech. To mitigate the difficulties of using seq2seq models to align long sequences, we down-sample the input spectral feature along the temporal dimension and equip the synthesis model with a discretized mixture of logistic (MoL) attention mechanism. Since the phoneme recognizer is trained with large speech recognition data corpus, the proposed approach can conduct any-to-many voice conversion. Objective and subjective evaluations show that the proposed any-to-many approach has superior voice conversion performance in terms of both naturalness and speaker similarity. Ablation studies are conducted to confirm the effectiveness of feature selection and model design strategies in the proposed approach. The proposed VC approach can readily be extended to support any-to-any VC (also known as one/few-shot VC), and achieve high performance according to objective and subjective evaluations.
\end{abstract}


%
\IEEEpeerreviewmaketitle

\section{Introduction}
\label{sec1:intro}
\IEEEPARstart{V}{oice} conversion (VC) aims to convert the non-linguistic information of a speech utterance while keeping the linguistic content unchanged. The non-linguistic information may refer to speaker identity, emotion, accent or pronunciation, to name a few. In this paper, we focus on the problem of speaker identity conversion.
Potential applications of VC techniques include entertainment, personalized text-to-speech, pronunciation or accent correction, etc.

Based on the number of source speakers and target speakers that a single VC system can support, we can categorize current VC approaches into one-to-one VC, many-to-one VC, many-to-many VC, any-to-many VC and any-to-any VC. 
Conventional VC approaches focus on one-to-one VC, which requires parallel training data between a pair of source-target speakers. At the training stage of the conventional VC pipeline, acoustic features are first extracted from the source and target utterances. The acoustic features of parallel utterances are then aligned frame-by-frame using alignment algorithms, such as dynamic time warping (DTW) \cite{bundy1984dynamic}. 
A conversion model is trained to learn the mapping function between time-aligned source and target acoustic features, which can be Gaussian mixture models (GMMs) \cite{stylianou1998continuous,toda2007voice}, artificial neural networks (ANNs) \cite{desai2010spectral,mohammadi2014voice,nakashika2015voice,sun2015voice}, etc.
These approaches perform frame-wise conversion on spectral features, i.e., the converted speech has the same duration as the source speech. This restricts the modeling of the speaking rate and duration. Recent studies show that the alignment phase can be surpassed through using sequence-to-sequence (seq2seq) model \cite{zhang2019sequence, tanaka2019atts2s} for direct source-target acoustic modeling, and this approach can achieve better VC performance.
Since one-to-one VC is limited to supporting only one particular pair of source and target speakers, VC researchers have explored many-to-one VC approaches to extend the versatility of VC approaches. Among these approaches, the one based on phonetic posteriorgrams (PPGs) is widely used \cite{sun2016phonetic, liu2018hccl, liu2019jointly}. PPGs are computed from an ASR acoustic model and are often assumed to be speaker-independent content representations. The many-to-one VC approaches concatenate a PPG extractor with a target-speaker dependent PPG-to-acoustic synthesis model. 
Many approaches have been proposed to further extend VC approaches to support many-to-many conversion. These techniques can be classified into two categories. 
The first category requires text supervision during training stage. This includes the PPG-based methods and the non-parallel seq2seq methods \cite{zhang2019non,kameoka2020many}. The second category does not require text supervision. This includes the model using auto-encoders \cite{qian2019autovc}, variational auto-encoders \cite{hsu2016voice}, generative adversarial networks \cite{gao2018voice, kaneko2017parallel, fang2018high, kameoka2018stargan} and their combinations \cite{hsu2017voice, chou2018multi, kameoka2018acvae}. 

This paper focuses on developing an any-to-many VC approach, which is expected to convert a source speech signal from an arbitrary speaker to a target speaker appearing in the training set. 
Few any-to-many VC approaches have been reported in the literature. Specific many-to-many VC approaches, which can directly be adopted for any-to-many conversion with text supervision during training stage, include the many-to-many PPG-based approaches and the non-parallel seq2seq approaches. These approaches have the presumption that the spoken content feature extractor/encoder in the many-to-many VC approaches can generalize well to source speakers that are unseen during training process.

\begin{figure}[t]
	\centering
	\includegraphics[width=8.8cm]{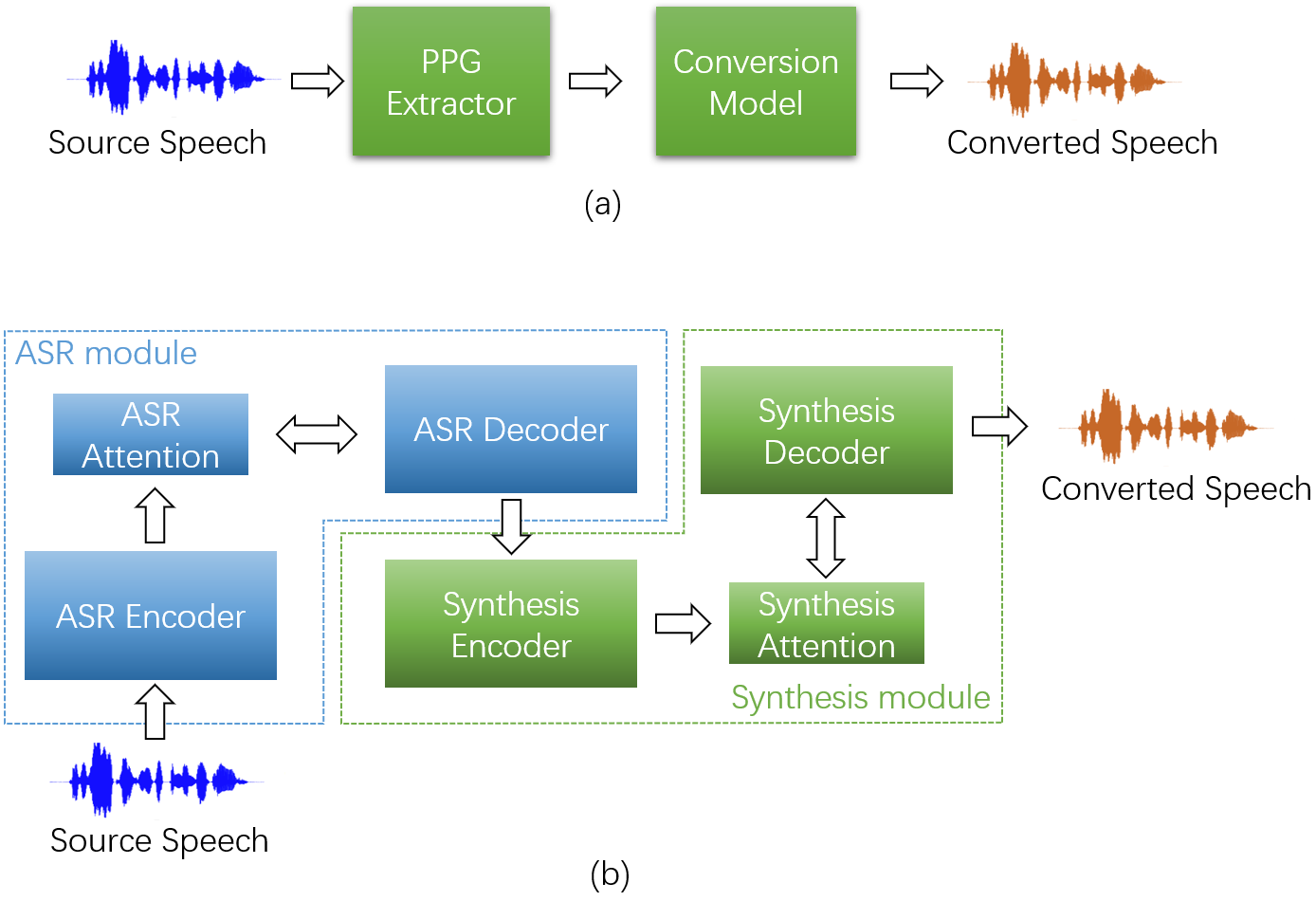}				
	\caption{Schematic diagram of (a) the PPG-based and (b) non-parallel seq2seq VC approaches.}
	\label{sec1:ppg_seq2seq}
\end{figure}

The use of PPGs in any-to-many VC concatenates the speaker-independent PPG extractor and a multi-speaker framewise feed-forward conversion model, as shown in Fig.~\ref{sec1:ppg_seq2seq}~(a). Such an approach has several deficiencies: 
First, the PPG model is usually trained with the HMM-GMM/DNN-based phonetic alignments with acoustic features. Since posterior probability values in PPGs are usually bimodal (with highest values close to one and other values close to zero), if the alignments are inaccurate, mis-pronunciations will often occur downstream in the VC pipeline. 
Second, the conversion model usually adopts a feed-forward neural network (e.g., a bidirectional LSTM model) which maps PPGs framewise to acoustic features. Conditioned on the input PPGs, these conversion models predict each acoustic frame independently. 
This is unfavorable in terms of VC performance (especially for scenarios where training data is sparse), because acoustic frames in an utterance are highly correlated.

To address the aforementioned deficiencies of PPG-based any-to-many VC approaches, this paper proposes the use of a different content feature extractor from the PPG model and a seq2seq auto-regressive synthesis model.
We first train an end-to-end hybrid connectionist-temporal-classification-attention (CTC-attention) phoneme recognizer, where the encoder has a bottle-neck layer. A bottle-neck feature extractor (BNE) is obtained from the phoneme recognizer and is used to extract bottle-neck features (BNF) as spoken content representations of speech signals.
To mitigate the difficulties of using a seq2seq model to align the content features and spectral features, we down-sample the input speech features along the temporal dimension by a factor of four. We then train a multi-speaker seq2seq BNF-to-spectral synthesis model, where each speaker is represented as an one-hot vector. To facilitate seq2seq modeling, the synthesis model is equipped with a mixture-of-logistic (MoL) location-relative attention module. We term this VC approach as BNE-Seq2seqMoL below and details are presented in Section~\ref{sec4}.

The synthesis module in the BNE-Seq2seqMoL apporach uses one-hot vectors to represent speaker identities. This paper also explores extending this any-to-many VC approach to support any-to-any conversion. The main idea is to incorporate speaker representations into the synthesis module with speaker embedding vectors which can generalize to unseen speakers. To achieve this, a speaker encoder is utilized to generate a fixed-dimensional speaker vector from a speech signal.

The contributions of this paper include: (1) We propose a novel MoL attention-based seq2seq model for any-to-many voice conversion. The proposed model can achieve high VC performance in terms of the naturalness of the generated speech, as well as its speaker similarity; but the model has a shorter system pipeline and a simpler training procedure than other analogous systems. (2) We present a straight-forward methodology to extend the any-to-many VC approach to support any-to-any conversion.


The rest of this paper is organized as follows: Section~\ref{sec2} reviews related work. Section~\ref{sec3} and~\ref{sec4} present a baseline model and the BNE-Seq2seqMoL approach, respectively. Experiments are described in Section~\ref{sec5} and Section~\ref{sec6} concludes this paper.

 
\section{Related work}
\label{sec2}

\subsection{Attention mechanisms in seq2seq models}
Sequence-to-sequence models equipped with attention mechanism have been very popular in VC and TTS tasks. 
The BNE-Seq2seqMoL approach proposed in this paper uses a location-relative attention mechanism, which is first introduced by Graves \cite{graves2013generating}. Inspired by \cite{vasquez2019melnet}, we incorporate a discretized mixture of logsitics (MoL) distribution \cite{salimans2017pixelcnn++} to model the attention weights in each decoding step in the BNE-Seq2seqMoL approach. Modifications are applied to make the alignment process strictly monotonic. 
Details are presented in Section~\ref{sec4:seq2seqmol}.

\subsection{A Seq2seq baseline model}
As described in Section~\ref{sec1:intro}, the non-parallel seq2seq approaches are another class of solutions for any-to-many VC. They usually cascade a seq2seq ASR model and a multi-speaker seq2seq synthesis model, as shown in Fig.~\ref{sec1:ppg_seq2seq}~(b). These approaches also have several drawbacks despite their strong sequence modeling ability: First, the pipeline is very long, which means that the model contains many parameters, resulting in a complicated and slow training process. Second, the ASR module usually adopts beam search algorithms to reduce recognition errors during inference. This slows down the conversion process. Third, the multi-speaker seq2seq synthesis model usually uses an attention module to align the hidden states of the encoder and decoder. There has been evidence of instability in this attention-based alignment procedure, as it may introduce missing or repeating words, incomplete synthesis, or an inability to generalize to longer utterances \cite{battenberg2020location}.
To make the synthesis more robust, our prior work \cite{liu2020transferring} incorporates a rhythm model, which guides the explicit temporal expansion process of the hidden representations output from the encoder. The auto-regressive decoder adopts a local attention mechanism within a small window, which further corrects possible alignment errors for high-fidelity synthesis. The prior work focuses on maintaining source speaking styles in the converted speech. The any-to-many conversion performance of this approach, however, needs to be thoroughly examined. In this paper, we present proper modifications in the prior approach \cite{liu2020transferring} and re-design a robust, non-parallel, seq2seq any-to-many VC approach. This pipeline concatenates a seq2seq phoneme recognizer (Seq2seqPR) and a multi-speaker duration informed attention network (DurIAN) for systhesis. This technique is referred to as Seq2seqPR-DurIAN below. Details are presented in Section~\ref{sec3}.

\subsection{Recognition-synthesis VC approaches}
Both the Seq2seqPR-DurIAN and BNE-Seq2seqMoL VC approaches proposed in this paper belong to the class of recognition-synthesis-based approaches, where an ASR module is built to extract spoken content representations and the synthesis module is used to predict acoustic features from the spoken content representations. Compared with the non-parallel seq2seq VC approach proposed in \cite{zhang2019non}, the Seq2seqPR-DurIAN approach utilizes a more robust synthesis module to mitigate possible attention alignment errors, where a duration model is incorporated to provide explicit phoneme-level alignment information. Besides, the Seq2seqPR-DurIAN approach has a simpler and more direct training procedure, whereas the approach in \cite{zhang2019non} uses complicated loss function and the training procedure alternates between a generation step and an adversarial step.
Specifically, the loss objective in [13] contains a phoneme sequence classification loss, an embedding contrastive loss, an adversarial speaker classification loss, an adversarial speaker mean squared error loss, a speaker encoder loss and an acoustic feature prediction loss.

A recent study, which is related to the BNE-Seq2seqMoL proposed in this paper, uses a pre-trained ASR encoder and a pre-trained TTS decoder to initialize parameters of the ultimate encoder-decoder-based VC model \cite{huang2020pretraining}. The ASR and TTS models are pre-trained with large-scale corpora with a multi-stage process, e.g., the ASR pre-training process contains TTS decoder pre-training, ASR encoder pre-training and ASR decoder pre-training. Then the VC model is fine-tuned on the pre-trained parameters with a small number of parallel utterances between a specific pair of source and target speakers. In comparison, the proposed BNE-Seq2seqMoL adopts a simplified two-stage training scheme, i.e., a seq2seq phoneme recognizer training stage and a multi-speaker MoL attention based seq2seq synthesis model training stage. After this two-stage training, the BNE-Seq2seqMoL approach can directly support any-to-many voice conversion.


\section{Seq2seqPR-DurIAN-based voice conversion}
\label{sec3}

The Seq2seqPR-DurIAN approach concatenates a seq2seq phoneme recognizer (Seq2seqPR) and a multi-speaker duration informed attention network (DurIAN). 
The Seq2seqPR model is adopted to predict an $L$-length phoneme sequence $Y=\{y_l \in \mathcal{U}|l=1,\cdots,L\}$ from spectral feature vectors $X$ of a speech signal, where $\mathcal{U}$ is a set of distinct phonemes. The DurIAN model is utilized to generate spectral feature vectors $\hat{X}$ from an input phoneme sequence $Y$, conditioned on the speaker representations $s$ to achieve multi-speaker synthesis. 

\subsection{Seq2seq phoneme recognizer}
\label{sec3:seq2seqPR}
We adopt the hybrid CTC-attention model structure for the seq2seq phoneme recognizer, which has similar network structure to \cite{kim2017joint}, as shown in Fig.~\ref{sec3:hybrid_ctc_att}.

\subsubsection{CTC and attention-based modeling}
CTC is a latent variable model that monotonically maps an input sequence to an output sequence of shorter length \cite{graves2006connectionist}. An additional ``blank" symbol is introduced into frame-wise phoneme sequence $Z=\{z_t\in \mathcal{U}\cup\text{blank}|t=1,\cdots,T\}$, where $T$ is the number of spectral frames. By using conditional independence assumptions, the posterior distribution $p(Y|X)$ is factorized as follows:
\begin{equation} \label{eq1}
P(Y|X) = \underbrace{\sum_{Z}\prod_{t}p(z_t|z_{t-1}, Y)p(z_t|X)}_{\triangleq p_{ctc}(Y|X)}p(Y)
\end{equation}

We define $p_{ctc}(Y|X)$ as the CTC objective function, where the frame-wise posterior distribution $p(z_t|X)$ is conditioned on all inputs X, and it is quite natural to be modeled with a deep neural network (e.g., LSTM model). The summation over $Z$ in Eq.~\ref{eq1} can be efficiently computed using a dynamic programming algorithm.

The attention-based approach directly estimates the posterior $p(Y|X)$ based on the probability chain rule as:

\begin{equation} \label{eq2}
P(Y|X) = \underbrace{\prod_{l}p(y_l|y_1,\cdots,y_{l-1}; X)}_{\triangleq p_{att}(Y|X)},
\end{equation}
where we define $p_{att}(Y|X)$ as an attention-based objective function, which can be conveniently modeled with an attention-based encoder-decoder model.

\subsubsection{Model structure and training objective}

\begin{figure}[t]
	\centering
	\includegraphics[width=4cm]{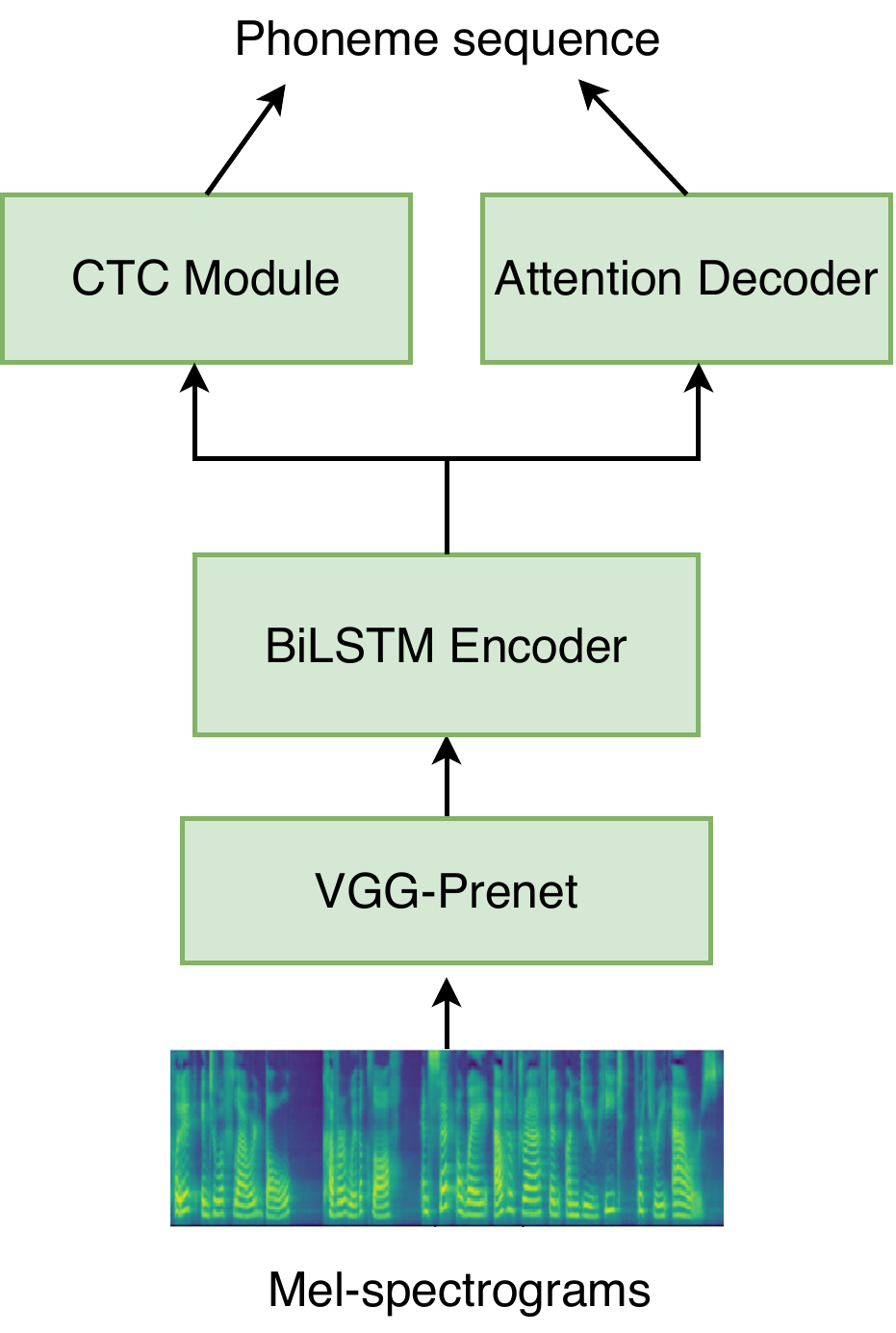}				
	\caption{Hybrid CTC-attention model structure \cite{kim2017joint} for phoneme recognition.}
	\label{sec3:hybrid_ctc_att}
\end{figure}

Following \cite{kim2017joint}, we regard the CTC objective as an auxiliary task to train the attention model encoder, which contains a VGG-Prenet and a bidirectional LSTM (BiLSTM) encoder, as shown in Fig.~\ref{sec3:hybrid_ctc_att}. 

The input spectral features $X$ are 80-dimensional log mel-spectrograms, on which we conduct utterance-level mean-variance normalization before feeding into the recognizer model. 
The VGG-Prenet sub-samples the input features by a factor of 4 in time scale using two VGG-like max pooling layers. Then the hidden feature maps from the VGG-Prenet are fed into the BiLSTM encoder, which contains 4 BiLSTM layers with 512 hidden units per direction. The CTC module has one fully-connected (FC) layer. The attention decoder uses location-sensitive attention and has one decoder LSTM layer with hidden size of 1024.

The training objective to be maximized is a logarithmic linear combination of the CTC and attention objectives, i.e., $p_{ctc}(Y|X)$ in Eq.~\ref{eq1} and ${p_{att}(Y|X)}$ in Eq.~\ref{eq2}:
\begin{equation} \label{eq3}
\mathcal{J}_{Seq2seqPR} = \lambda\log P_{ctc}(Y|X) + (1-\lambda)\log P_{att}(Y|X)
\end{equation}
where $\lambda \in [0,1]$ is a hyper-parameter weighting the CTC objective and the attention objective. In this paper, we set $\lambda$ to be $0.5$.

\subsection{DurIAN synthesis model}
\label{sec3:DurIAN}

\begin{figure}[t]
	\centering
	\includegraphics[width=9cm]{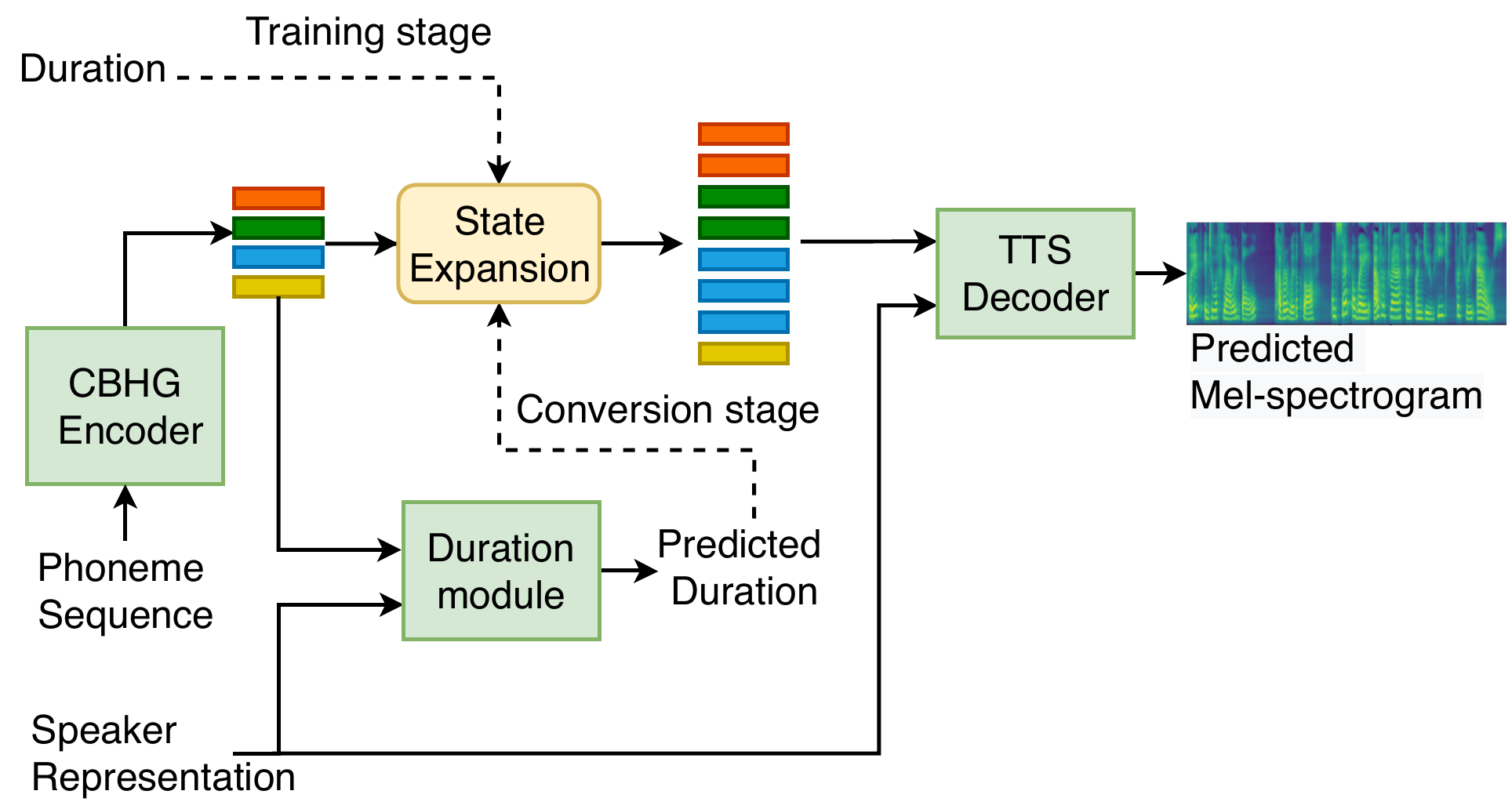}				
	\caption{Duration informed attention network (DurIAN) used in the Seq2seqPR-DurIAN VC approach.}
	\label{sec3:durian_tts}
\end{figure}

The DurIAN synthesis model used in this paper is inspired by \cite{yu2019durian}, which is trained to predict the mel-spectrogram $X$ from an input phoneme sequence $Y$, as shown in Fig.~\ref{sec3:durian_tts}. 
Attention-based seq2seq TTS models such as Tacotron are error prone in the alignment procedure, which leads to missing or repeating words, incomplete synthesis or an inability to generalize to longer utterances. To address this issue, we incorporate a duration module into the synthesis model. A similar idea has been used in \cite{ren2019fastspeech}.

A CBHG encoder \cite{wang2017tacotron} is adopted to transform phoneme sequences into hidden representations. In the state expansion procedure, the hidden representations are expanded by repeating along the temporal axis according to the provided phoneme-level durational information, such that the expanded representations have the same number of frames as the spectral features. 
An auto-regressive RNN-based TTS decoder is used to generate mel-spectrograms from the expanded spectral features, conditioned on the speaker representations (e.g., one-hot vectors) to support multi-speaker generation. In the any-to-many VC setting, the speaker identity is represented with one-hot vectors. A speaker embedding table is jointly optimized with the remaining parts of the DurIAN model. Speaker embedding vectors are appended to every frames of the expanded encoder hidden representations. 
Note that in the any-to-any VC setting, speaker vectors generated from a pre-trained speaker encoder are used to represent speaker identity. The details are presented in Section III D.
The TTS decoder has similar network structure as the one in Tacotron 1 \cite{wang2017tacotron}. The only difference is that the attention context concatenated with the decoder prenet output is replaced with the corresponding encoder state in the expanded hidden representations. Similar to Tacotron 1, we make the decoder generate $r$ non-overlapped mel-spectrogram frames at each decoding step to accelerate the training and synthesis. 

The duration module employs an RNN-based model which consists of three BiLSTM layers. The input to the duration module contains the un-expanded hidden states from the CBHG encoder and speaker identity representation. As in the TTS decoder, in the any-to-many VC setting, speakers are presented with one-hot vectors and a speaker embedding table is jointly learned with the duration module. In the any-to-any tasks, speaker vectors from the same speaker encoder are adopted to represent speaker identity. Speaker vectors are appended to all frames of the CBHG encoder output.

\subsection{Extension to support any-to-any conversion}
\label{sec3:a2a}
To extend the Seq2seqPR-DurIAN to support any-to-any conversion, we use an additional speaker encoder model to generate the speaker vector, which is used to condition the DurIAN synthesis module to generate speech with the identity of an arbitrary target speaker. 
The speaker encoder takes acoustic vector sequence with various number of frames computed from a speech signal and outputs a fixed-dimensional speaker embedding vector.
The DurIAN model uses the speaker embedding vector computed from a desired target speaker as auxiliary conditioning to control the vocal identity of the generated speech.
As in \cite{wan2018generalized, liu2020accent}, we train the speaker encoder to optimize a generalized end-to-end (GE2E) speaker verification loss. Embeddings of utterances from the same speaker are expected to have high cosine similarity, while those from different speakers are distant.
The speaker encoder adopts an LSTM-based model structure, which has 3 layers with 256 hidden nodes, followed by a projection layer of 256 units. L2-normalized hidden state of the last layer is regarded as the speaker embedding vector.

\section{BNE-Seq2seqMoL-based voice conversion}
\label{sec4}
The proposed BNE-Seq2seqMoL approach combines a bottle-neck feature extractor (BNE) with a multi-speaker  mixture of logistic (MoL) attention-based seq2seq synthesis model. The BNE is used to compute dense and rich content features from mel-spectrograms, while the MoL attention-based seq2seq model (Seq2seqMoL) is adopted to generate mel-spectrograms auto-regressively. Details of the BNE and the Seq2seqMoL model are presented in Section~\ref{sec4:bnf} and \ref{sec4:seq2seqmol} respectively. The conversion procedure and extension to any-to-any VC are presented in the later parts of this section.

\subsection{Bottle-neck feature extractor}
\label{sec4:bnf}
\begin{figure}[t]
	\centering
	\includegraphics[width=6cm]{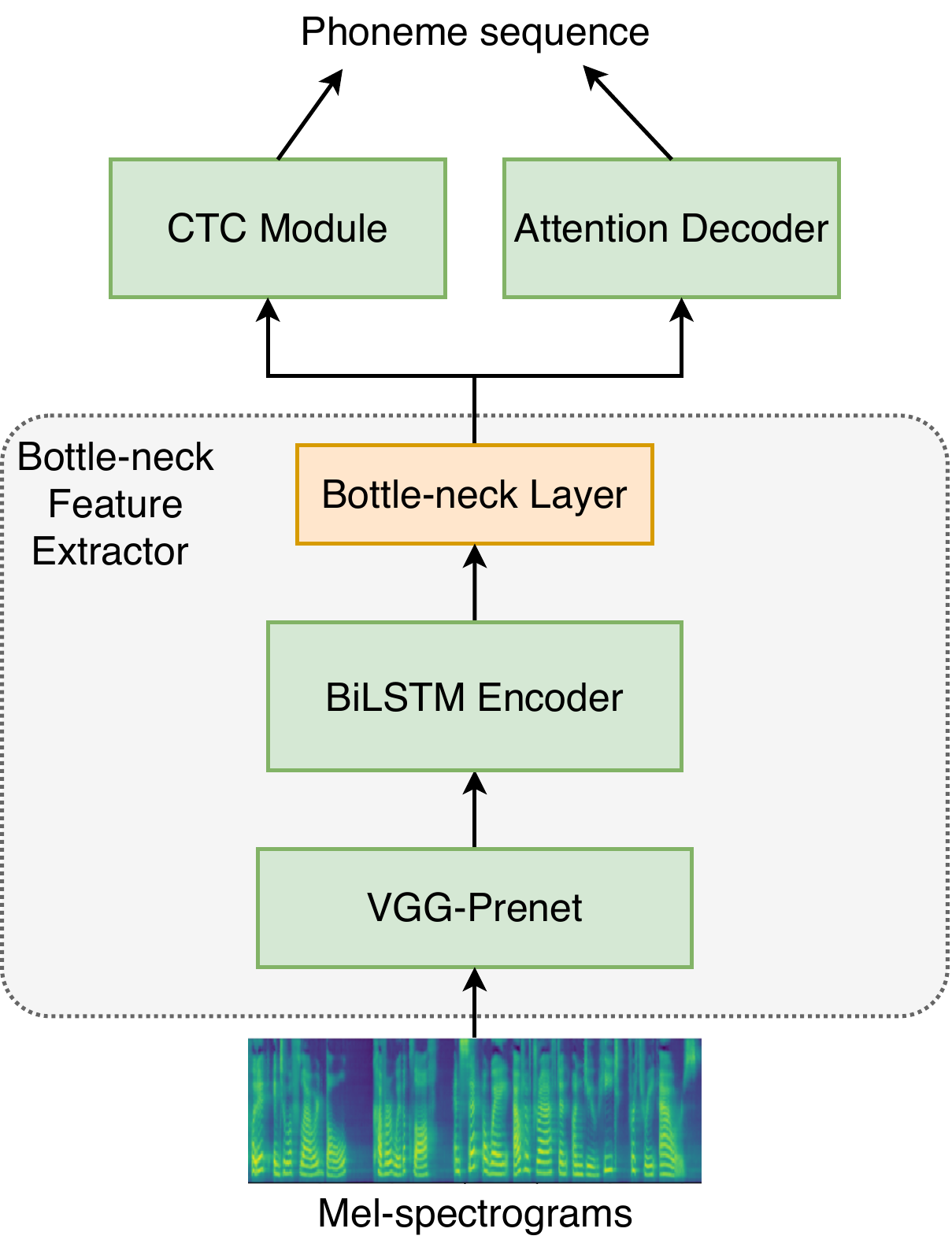}
	\caption{Illustration of the training process of the bottleneck feature extractor (BNE) used in the proposed BNE-Seq2seqPR VC approach.}
	\label{sec4:hybrid_att_ctc_asr_bnf}
\end{figure}

We obtain a bottle-neck feature extractor from an end-to-end hybrid CTC-attention phoneme recognizer. The phoneme recognizer has the same network structure as the one introduced in Section~\ref{sec3:seq2seqPR}, except that we incorporate an additional bottle-neck layer into the recognizer, as illustrated in Fig.~\ref{sec4:hybrid_att_ctc_asr_bnf}. The bottle-neck layer is a fully-connected layer with hidden size of 256.
The training objective is the same as that used in Section~\ref{sec3:seq2seqPR} (See~Eq.~\ref{eq3}). After training, we drop the CTC module and attention decoder from the phoneme recognizer and use the remaining part as the bottle-neck feature extractor. 
The bottle-neck features computed from speech signals are regarded as spoken content representations and are presumed to be speaker-invariant \cite{adi2019reverse}.

\subsection{Seq2seqMoL synthesis}
\label{sec4:seq2seqmol}
\begin{figure}[t]
	\centering
	\includegraphics[width=8cm]{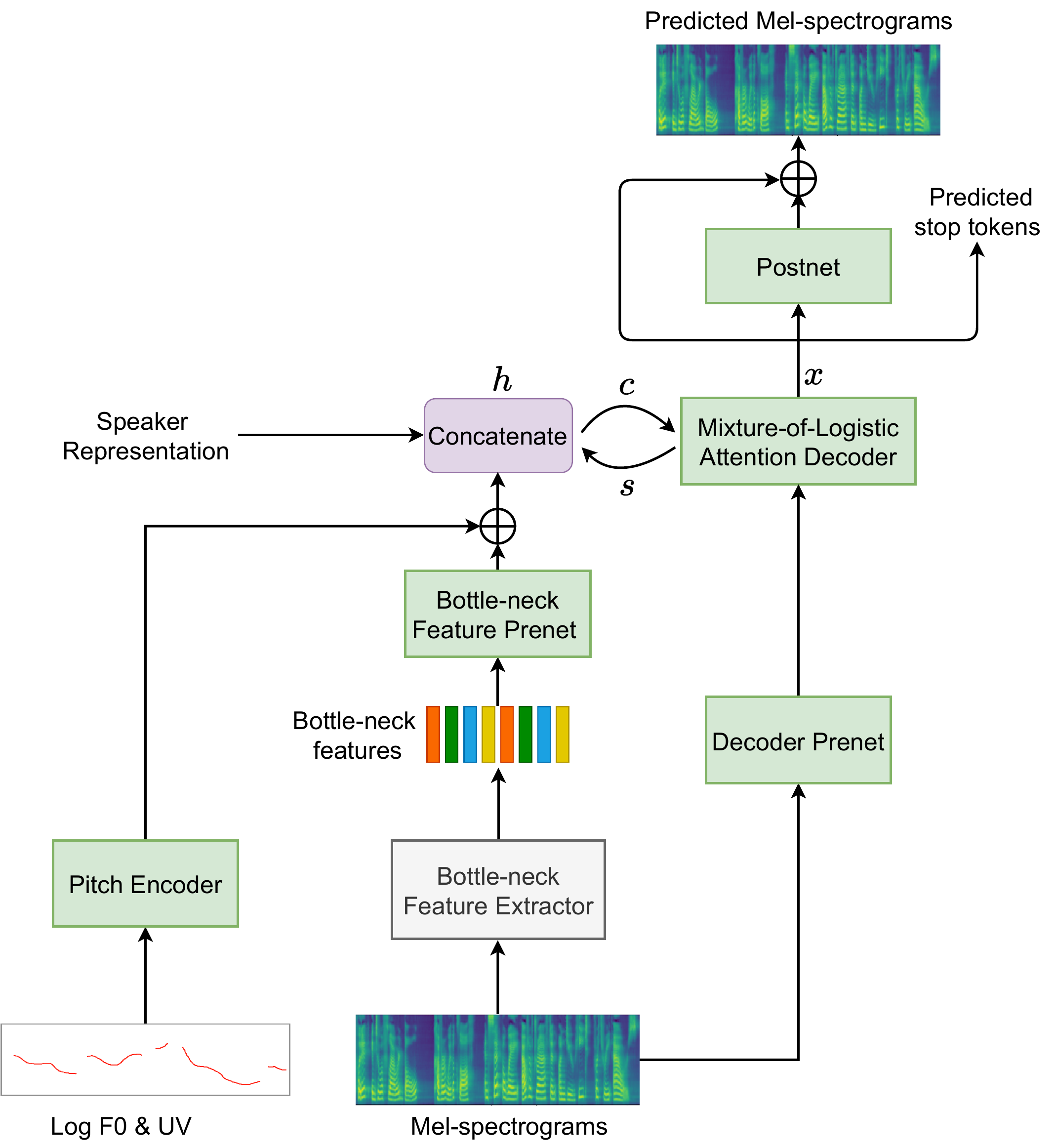}				
	\caption{Schematic illustration of the training stage of the synthesis model in the proposed BNE-Seq2seqMoL approach.}
	\label{sec4:seq2seq_bnf2mel}
	\vspace{-20pt}
\end{figure}

The training procedure of the seq2seq based synthesis model is depicted in Fig.~\ref{sec4:seq2seq_bnf2mel}. A well-trained BNF extractor presented in Section~\ref{sec4:bnf} is adopted as an off-line content feature extractor. The synthesis model can be regarded as an encoder-decoder model, where the encoder contains two simple networks, i.e., a bottle-neck feature prenet and a pitch encoder.

\subsubsection{Bottle-neck feature prenet and pitch encoder}
The bottle-neck feature prenet contains two bidirectional GRU layers, which have 256 hidden units per direction. The pitch encoder employs convolution network structure, which takes continuously interpolated logarithmic F0 (Log-F0) and unvoiced-voiced flags (UV) features as input. Log-F0s and UVs are computed with the same frame-shift as the one used to extract mel-spectrograms. Since the bottle-neck feature extractor down-samples the mel-spectrograms by a factor of 4 along the time axis, the bottle-neck features only have a quarter of the frames in the corresponding Log-F0s or UVs. To make BNFs, Log-F0s and UVs have the same time resolution, we also down-sample Log-F0 and UVs by a factor of 4 along time axis. This is achieved by using two 1-dimensional convolution layers with a stride of 2, where the hidden-dimension is 256. To remove possible speaker information, we add an instance normalization layer without affine transformation after each convolution layer in the pitch encoder.

The outputs of the pitch encoder and the bottle-neck feature prenet are added element-wise. In any-to-many conversion, one-hot vectors are used as the speaker representation, and an additional speaker embedding table is jointly trained with the whole synthesis network. Speaker vectors are concatenated to every frame of the encoder output.

\subsubsection{MoL attention-based decoder}
Decoder of the synthesis model adopts a similar auto-regressive network structure as the one used in Tacotron 2, except that a location-relative discretized mixture of logistics (MoL) attention mechanism is used.

\begin{figure}[t]
	\centering
	\includegraphics[width=9cm]{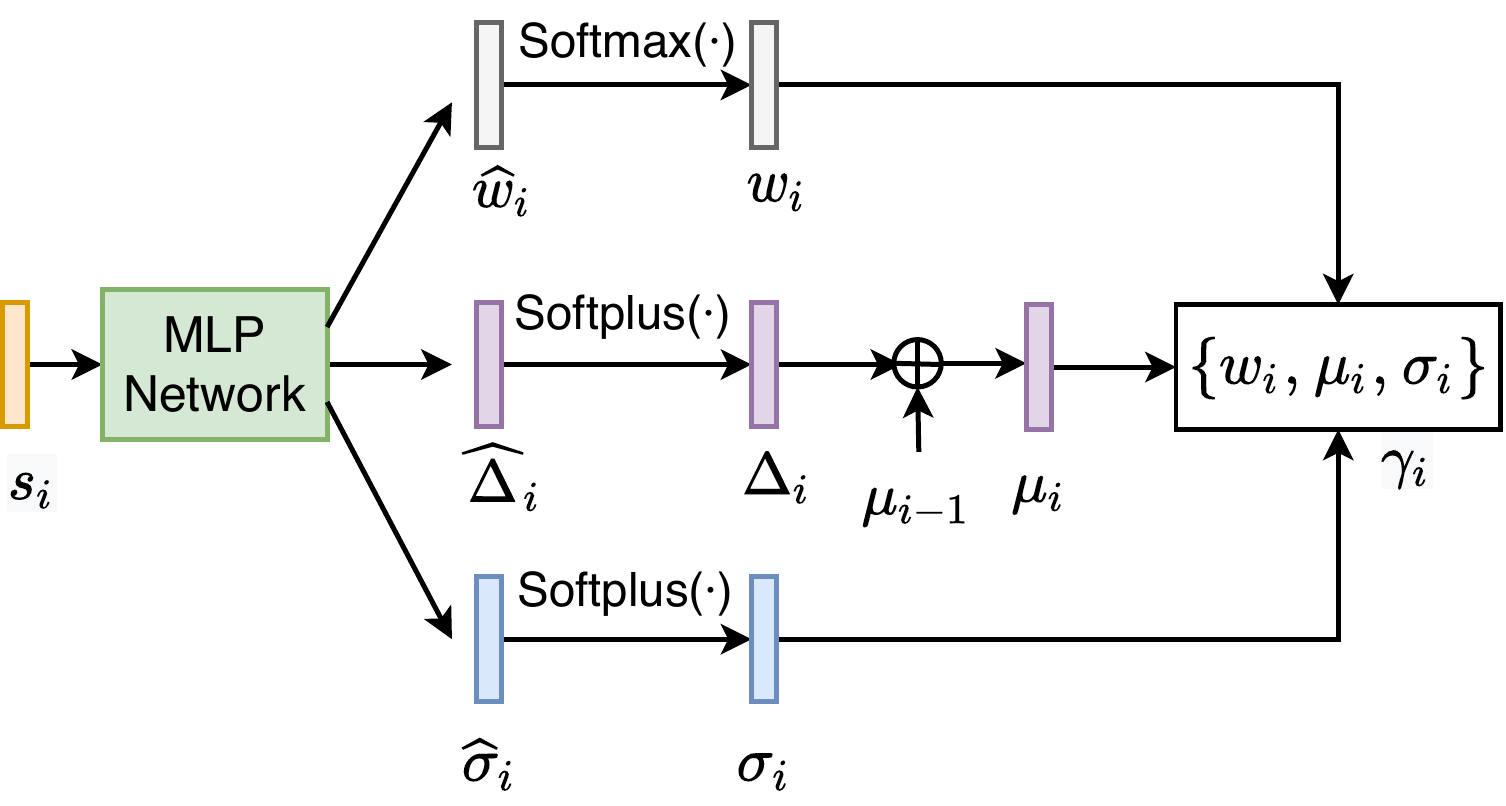}				
	\caption{Computation procedure of the mixture of logistic distribution parameters $\gamma_i$  for attention weights in decoder step $i$ in the proposed BNE-Seq2seqMoL VC approach.}
	\label{sec4:mol_att}
\end{figure}

Let us denote the encoder outputs as $\{h_j\}_{j=1}^{\tilde{T}}$, where $\tilde{T} = \frac{1}{4}T$ and $T$ is the number of frames in the mel-spectrograms. An attention RNN (Eq.~\ref{eq4}) produces hidden state $s_i$ at decoder step $i$. Then the attention mechanism consumes $s_i$ to produce the alignment $\alpha_i\in\mathbf{R}^{\tilde{T}}$ (Eq.~5). The context vector, $c_i$, which is fed to the decoder RNN, is computed using the alignment $\alpha_i$ to produce a weighted average of encoder states $\{h_j\}_{j=1}^{\tilde{T}}$ (Eq.~6). The decoder RNN takes $s_i$ and $c_i$ as input, whose output $d_i$ is used together with the context vector $c_i$ to produce mel-spectrogram frames at the current decoder step by a linear layer (Eq.~7 and 8). 
\begin{align}\label{eq4}
s_i &= \text{RNN}_{\text{Att}}([x_{i-1}, c_{i-1}], s_{i-1}) \\
\alpha_{i} &= \text{Attention}(s_i) \\
c_i &= \sum_{j=1}^{\tilde{T}}\alpha_{i,j}h_j \\
d_i &= \text{RNN}_{\text{Dec}}([c_i, s_i], d_{i-1}) \\
x_i &= \text{Linear}_{\text{Out}}(d_i, c_i)
\end{align}
The attention mechanism is similar to the one used in \cite{vasquez2019melnet}, which is a location-relative extension from a purely location-based mechanism proposed in \cite{graves2013generating}. The attention alignment weights correspond to a learned attention distribution $\phi(\cdot;\gamma_i)$, where $\gamma_i$ is the distribution parameters computed using a simple multi-layer perception (MLP) network from the attention RNN state $s_i$. We use a discretized MoL \cite{salimans2017pixelcnn++} for the attention distribution $\phi_i(\cdot;\gamma_i)$. At each decoder step, a set of distribution parameters $\gamma_i = \{w_i^k, \mu_i^k, \sigma_i^k \}_{k=1}^K$ is computed, corresponding to $K$ mixture coefficients, means and scales. In this paper, the number of mixtures are set to be 5. The computation procedure is shown as below and also illustrated in Fig.~\ref{sec4:mol_att}.
\begin{align} \label{eq5}
&(\hat{w}_i,\hat{\Delta}_i,\hat{\sigma_i}) = \text{MLP}(s_i)
\end{align}
\begin{align}
&w_i = \text{SM}(\hat{w}_i),\quad \Delta_i=\text{SP}(\hat{\Delta}_i),\quad \sigma = \text{SP}(\hat{\sigma}_i)
\end{align}
\begin{align}
&\mu_i = \mu_{i-1} + \Delta_i
\end{align}
where $\text{SM}(\cdot)$ represents the softmax function and $\text{SP}(\cdot)$ represents the softplus function.
Note that in Eq. 11, we add a positive shift to the component mean in the previous decoding step, which makes the center of each component logistic distribution moving towards to the end of the input sequence in a monotonic way.
Given the computed MoL distribution parameters $\gamma_i$, the attention weight $\alpha_{i,j}$ is obtained from the discretized attention distribution $\phi_i(\cdot;\gamma_i)$ at decoder step $i$ as:
\begin{equation} \label{eq6}
\begin{split}
&\alpha_{i,j} = \phi_i(j;\gamma_i) \\ 
&= \sum_{k=1}^{K}w_i^k[\sigma(\frac{j + 0.5 - \mu_i^k}{\sigma_i}) - \sigma(\frac{j - 0.5 - \mu_i^k}{\sigma_i})] 
\end{split}
\end{equation} 
where $\sigma$ denotes the sigmoid function.

Following Tacotron 1 and 2, a decoder prenet containing two linear layers and a residual convolution based postnet are added to the synthesis decoder. We also let the decoder predict stop tokens, which are used to stop the decoding procedure when the stopping probability reach a threshold $0.5$.

The training objective is to minimize the MSE loss between the ground truth mel-spectrogram $X$ and the predicted $\hat{X}$, in combination with a binary cross-entropy loss on the stop token predictions.

\subsection{Conversion procedure}
Given a speech utterance from an arbitrary source speaker, the approach first computes the mel-spectrogram, continuous Log-F0s and UV flags. Then the BNF extractor is used to extract content features from the mel-spectrogram. Log-F0s are converted linearly in log-scale from the source to target using log-scaled F0 statistics of the source and target speakers, as:

\begin{equation}\label{eq7}
 \text{Log-F0}_{\text{vc}} = \frac{\sigma_{\text{target}}}{\sigma_{\text{source}}}(\text{Log-F0}_{\text{source}} - \mu_{\text{source}}) +  \mu_{\text{target}}
\end{equation}
where $\mu$'s and $\sigma$'s represent the mean and standard deviation of the log-scaled F0.

The bottle-neck features, $\text{Log-F0}_{\text{vc}}$ and UV flags are added element-wise after going through the bottle-neck features prenet and the pitch encoder respectively. The output is concatenated with the target speaker embedding vector to form the encoder outputs. The MoL attention decoder then generates the converted mel-spectrogram from the encoder outputs in an auto-regressive manner. A neural vocoder is finally used to generate waveform from the converted mel-spectrogram.

\subsection{Extension to any-to-any conversion}
\label{sec4:a2a}
We use the same speaker encoder model introduced in Section~\ref{sec3:a2a} to generate the speaker vector for an arbitrary target speaker. We replace the one-hot speaker representation with the speaker-encoder-generated speaker vector in the BNE-Seq2seqMol any-to-many approach, such that it supports any-to-any conversion. Details of the speaker encoder are the same as that presented in Section~\ref{sec3:a2a}.

\section{Experiments}
\label{sec5}

\subsection{Datasets}
\label{sec5:dataset}
The datasets used in this paper are all publicly available. LibriSpeech (960 hours) \cite{panayotov2015librispeech} is used to train the phoneme recognizer introduced in section~\ref{sec3:seq2seqPR} and \ref{sec4:bnf}. The LibriSpeech lexicon \footnote{\url{http://www.openslr.org/11/}} is used to obtain phoneme sequences from text transcripts.

In any-to-many voice conversion, we use the VCTK corpus \cite{veaux2017cstr} and the CMU ARCTIC database \cite{kominek2003cmu}. The VCTK corpus contains 44 hours of clean speech from 109 speakers. In this paper, we only use data from 105 VTCK speakers. We choose 600 utterances for validation set and another 600 utterances for test set, while the remaining utterances are for train set. The CMU ARCTIC database contains 1132 parallel recordings of English speakers. Data from four speakers are used: two female (clb and slt) and two male (bdl and rms). 
We choose 50 utterances for validation and another 50 utterances for testing. We randomly choose non-overlapped 250 utterances from the remaining utterances for each of the four speakers respectively, such that they do not have parallel utterances during training.

For any-to-any voice conversion, we use LibriSpeech (train-other-500), VoxCeleb1 \cite{Nagrani17} and VoxCeleb2 \cite{Chung18b} datasets to train the speaker encoder introduced in Section~\ref{sec3:a2a} and Seciton~\ref{sec4:a2a}. In total, there are more than 8K speakers, such that we expect the speaker encoder can generalize to any unseen speaker.
LibriTTS (train-clean-100 and train-clean-360) dataset \cite{zen2019libritts} together with the training set of VCTK corpus is used to train the synthesis models of the Seq2seqPR-DurIAN and BNE-Seq2seqMoL approaches. The CMU ARCTIC database is used only for the conversion stage in this setting. That is, the four speakers (bdl, clb, rms and slt) are all unseen during the training procedure. This simulates voice conversion from an arbitrary source speaker to an arbitrary target speaker, which forms a pilot version of any-to-any conversion.   

\subsection{Features and neural vocoder model}
Speech signals used in this paper are all re-sampled to 16kHz if the original sampling rate is different. Spectral features are all 80-dimensional log mel-spectrograms except that the speaker encoder takes 40-dimensional log mel-spectrograms as input. The 80-dimensional log mel-spectrograms are computed using 50ms Hanning window and 10ms frame shift, while the 40-dimensional ones are computed using 25ms Hanning window and 10ms frame shift. We use the PyWorld toolkit \footnote{\url{https://github.com/JeremyCCHsu/Python-Wrapper-for-World-Vocoder}} to extract F0s from speech signals and Log-F0s are obtained by taking logarithm on the linearly interpolated F0s. 
We use the open-source Montreal-forced-aligner (MFA) \cite{mcauliffe2017montreal} to obtain the phoneme-level durational information when training the synthesis model in the Seq2seqPR-DurIAN approach.

In this paper, the WaveRNN network \cite{kalchbrenner2018efficient} is used as the neural vocoder. The speech waveform is $\mu$-law quantized into 512-way categorical distributions. The open-sourced Pytorch implementation is used.\footnote{\url{https://github.com/fatchord/WaveRNN}} Since the mel-spectrograms capture all of the relevant details needed for high quality speech synthesis, we simply use ground-truth mel-spectrograms from multiple speakers to train the WaveRNN, without adding any speaker identity representations. We only use the VCTK training set to train the WaveRNN model.

\begin{table*}[ht]
\centering
\caption{Objective evaluation results for any-to-many voice conversion.}
\resizebox{\textwidth}{!}{\begin{tabular}{c||c|c|c|c||c|c|c|c||c|c|c|c||c|c|c|c}
\thickhline
\multirow{2}{*}{\begin{tabular}[c]{@{}c@{}}Conversion \\ pair\end{tabular}} &
  \multicolumn{4}{c||}{PPG-VC} &
  \multicolumn{4}{c||}{NonParaSeq2seq-VC} &
  \multicolumn{4}{c||}{Seq2seqPR-DurIAN} &
  \multicolumn{4}{c}{BNE-Seq2seqMoL} \\ \cline{2-17} 
 &
  MCD &
  \begin{tabular}[c]{@{}c@{}}F0\\ RMSE\end{tabular} &
  CER &
  WER &
  MCD &
  \begin{tabular}[c]{@{}c@{}}F0\\ RMSE\end{tabular} &
  CER &
  WER &
  MCD &
  \begin{tabular}[c]{@{}c@{}}F0\\ RMSE\end{tabular} &
  CER &
  WER &
  MCD &
  \begin{tabular}[c]{@{}c@{}}F0\\ RMSE\end{tabular} &
  CER &
  WER \\ \hline
F-M & 6.90 & 48.22 & 5.75 & 8.93 & 7.37 & 49.94 & 17.78 & 26.54 & 7.03 & 52.60 & 7.03 & 9.89 & \textbf{6.87}  & \textbf{44.55} & \textbf{3.00}  & \textbf{4.29}          \\ \hline
F-F     & 7.03  & 48.22 & 5.54 & 8.46 & 7.36 & \textbf{45.43} & 8.14  & 13.46 & 7.04 & 47.25 & 8.76 & 13.39 & \textbf{6.99} & 45.44 & \textbf{5.12} & \textbf{6.97}          \\ \hline
M-M & 6.96  & 51.4  & 5.19 & 7.50 & 7.36 & 53.86 & 24.78 & 39.80 & 7.14 & 51.89 & 7.33 & 14.16 & \textbf{6.94} & \textbf{50.37} & \textbf{4.72} & \textbf{7.22}          \\ \hline
M-F     & 7.25  & 56.36 & 5.32 & 7.80 & 7.39 & 50.37 & 19.05 & 25.71 & \textbf{7.09} & \textbf{45.09} & 8.32 & 14.72 & 7.18 & 56.31 & \textbf{3.61} & \textbf{5.76}      \\ \hline
Average & 7.04 & 51.05 & 5.45 & 8.17 & 7.37 & 49.90 & 17.44 & 26.38 & 7.08 & 49.21 & 7.86 & 13.04 & \textbf{6.99} & \textbf{49.17} & \textbf{4.11} & \textbf{6.06} \\ \thickhline
\end{tabular}}
\label{obj_table1}
\end{table*}

\begin{table*}[ht]
\centering
\caption{Objective evaluation results for any-to-any voice conversion.}
\resizebox{\textwidth}{!}{\begin{tabular}{c||c|c|c|c||c|c|c|c||c|c|c|c||c|c|c|c}
\thickhline
\multirow{2}{*}{\begin{tabular}[c]{@{}c@{}}Conversion \\ pair\end{tabular}} &
  \multicolumn{4}{c||}{PPG-VC} &
  \multicolumn{4}{c||}{NonParaSeq2seq-VC} &
  \multicolumn{4}{c||}{Seq2seqPR-DurIAN} &
  \multicolumn{4}{c}{BNE-Seq2seqMoL} \\ \cline{2-17} 
 &
  \multicolumn{1}{c|}{MCD} &
  \multicolumn{1}{c|}{\begin{tabular}[c]{@{}c@{}}F0\\ RMSE\end{tabular}} &
  \multicolumn{1}{c|}{CER} &
  \multicolumn{1}{c||}{WER} &
  \multicolumn{1}{c|}{MCD} &
  \multicolumn{1}{c|}{\begin{tabular}[c]{@{}c@{}}F0\\ RMSE\end{tabular}} &
  \multicolumn{1}{c|}{CER} &
  \multicolumn{1}{c||}{WER} &
  \multicolumn{1}{c|}{MCD} &
  \multicolumn{1}{c|}{\begin{tabular}[c]{@{}c@{}}F0\\ RMSE\end{tabular}} &
  \multicolumn{1}{c|}{CER} &
  \multicolumn{1}{c||}{WER} &
  \multicolumn{1}{c|}{MCD} &
  \multicolumn{1}{c|}{\begin{tabular}[c]{@{}c@{}}F0\\ RMSE\end{tabular}} &
  \multicolumn{1}{c|}{CER} &
  \multicolumn{1}{c}{WER} \\ \hline
F-M & 7.51 & 48.8 & \textbf{4.21} & \textbf{6.99} & 8.01 & 67.2 & 5.74 & 8.6 & 7.51 & 58.44 & 5.45 & 8.08 & \textbf{7.44} & \textbf{44.57} & 5.74 & 7.42 \\ \hline
F-F & \textbf{7.52} & 49.86 & 4.67 & 7.22 & 7.58 & \textbf{47.01} & 4.56 & \textbf{6.32} & 7.83 & 63.00 & 9.79 & 6.40 & 7.71 & 48.45 & \textbf{3.98} & 7.03 \\ \hline
M-M & 7.64 & 58.37 & \textbf{5.05} & \textbf{6.73} & 8.16 & 59.13 & 5.72 & 8.80 & \textbf{7.50} & 58.34 & 5.97 & 11.89 & 7.53 & \textbf{50.07} & 5.28 & 7.86 \\ \hline
M-F & \textbf{7.80} & 69.76 & \textbf{4.42} & \textbf{6.27} & 8.46 & \textbf{58.38} & 6.55 & 9.95 & 7.86 & 68.65 & 6.08 & 10.42 & 7.93 & 61.50 & 6.05 & 8.11 \\ \hline
Average & \textbf{7.62} & 56.70 & \textbf{4.59} & \textbf{6.80} & 8.05 & 57.93 & 5.64 & 8.42 & 7.68 & 62.11 & 6.82 & 9.20 & 7.65 & \textbf{51.15} & 5.26 & 7.60 \\ \thickhline
\end{tabular}
}
\label{obj_table2}
\end{table*}

\begin{table*}[ht]
\centering
\caption{Subjective evaluation results for four VC approaches: PPG-VC, NonParaSeq2seq-VC, Seq2seqPR-DurIAN and BNE-Seq2seqMoL ($95\%$ confidence intervals).}
\resizebox{\textwidth}{!}{
\begin{tabular}{c||c|c||c|c||c|c||c|c}
\thickhline
\multirow{2}{*}{\begin{tabular}[c]{@{}c@{}}Conversion\\ pair\end{tabular}} &
  \multicolumn{2}{c||}{PPG-VC} &
  \multicolumn{2}{c||}{NonParaSeq2seq-VC} &
  \multicolumn{2}{c||}{Seq2seqPR-DurIAN} &
  \multicolumn{2}{c}{BNE-Seq2seqMoL} \\ \cline{2-9} 
          & Naturalness & Similarity & Naturalness & Similarity & Naturalness & Similarity & Naturalness    & Similarity    \\ \hline
F-M & 2.23 $\pm$ 0.07 & 2.57 $\pm$ 0.06 & 2.95 $\pm$ 0.05 & 3.40 $\pm$ 0.06 & 3.43 $\pm$ 0.08 & 3.78 $\pm$ 0.06 & \textbf{3.72 $\pm$ 0.06} & \textbf{3.88 $\pm$ 0.05} \\ \hline
F-F & 2.78 $\pm$ 0.04 & 3.16 $\pm$ 0.07 & 3.17 $\pm$ 0.06 & 3.42 $\pm$ 0.08 & 3.53 $\pm$ 0.07 & 3.74 $\pm$ 0.09 & \textbf{3.76 $\pm$ 0.10} & \textbf{4.20 $\pm$ 0.11} \\ \hline
M-M & 2.35 $\pm$ 0.09 & 2.83 $\pm$ 0.10 & 2.86 $\pm$ 0.07 & 3.22 $\pm$ 0.08 & 3.27 $\pm$ 0.07 & 3.52 $\pm$ 0.08 & \textbf{3.84 $\pm$ 0.09} & \textbf{4.14 $\pm$ 0.08} \\ \hline
M-F & 2.56 $\pm$ 0.09 & 2.52 $\pm$ 0.09 & 3.34 $\pm$ 0.06 & 3.69 $\pm$ 0.09 & \textbf{3.72 $\pm$ 0.05} & \textbf{3.82 $\pm$ 0.04} & 3.66 $\pm$ 0.08 & 3.04 $\pm$ 0.09 \\ \hline
Average & 2.48 $\pm$ 0.08 & 2.78 $\pm$ 0.08 & 3.08 $\pm$ 0.06 & 3.43 $\pm$ 0.08 & 3.49 $\pm$ 0.07 & 3.72 $\pm$ 0.07 & \textbf{3.75 $\pm$ 0.08} & \textbf{3.82 $\pm$ 0.09} \\ \hline
Recording & 4.78 $\pm$ 0.07 & - & - & - & - & - & - & - \\ 
\thickhline
\end{tabular}
}
\label{sub_table1}
\end{table*}

\begin{table*}[]
\centering
\caption{Subjective evaluation results of the proposed BNE-Seq2seqMoL approach for any-to-any voice conversion ($95\%$ confidence intervals).}
\begin{tabular}{c|c|c|c|c|c}
\thickhline
\begin{tabular}[c]{@{}c@{}}
Conversion\\ pair\end{tabular} & F-M & M-M & F-F & M-F & Average \\ \hline
Naturalness & 3.42  $\pm$ 0.06 & 3.61  $\pm$ 0.09 & 3.70  $\pm$ 0.07 & 3.44  $\pm$ 0.05 & 3.54  $\pm$ 0.07 \\ \hline
Similarity & 3.53  $\pm$ 0.08 & 3.57  $\pm$ 0.06 & 3.75  $\pm$ 0.09 & 2.85  $\pm$ 0.10 & 3.43  $\pm$ 0.08 \\
\thickhline
\end{tabular}
\label{sub_table2}
\end{table*}

\subsection{Comparisons}
Four VC approaches are compared with experiments for any-to-many conversion, as well as extending to any-to-any conversion. We compare the proposed Seq2seqPR-DurIAN and BNE-Seq2seqMoL approaches with another two recently proposed approaches, namely, the PPG-based VC and the non-parallel seq2seq VC, as introduced in Section~\ref{sec1:intro}. The details of their implementation are presented below.

\textbf{PPG-VC}: This baseline approach has a network architecture similar to the N10 system \cite{liu2018wavenet} in VCC2018 \cite{lorenzo2018voice}. As shown in Fig.~\ref{sec1:ppg_seq2seq}~a, this approach consists of a PPG extractor and a multi-speaker conversion model. The PPG extractor adopts an RNN-based model structure, which contains 5 bidirectional gated recurrent unit (GRU) layers with 512 hidden units per direction. The multi-speaker conversion model also has an RNN structure, which consists of 4 bidirectional LSTM layers with 256 hidden units per direction. In any-to-many conversion, the speaker identity is presented as an one-hot vector and an additional speaker embedding table is learned together with other parts of the conversion model. The speaker embedding vector has a size of 256, and is concatenated with PPGs frame-by-frame. The conversion model is trained with the VCTK and CMU ARCTIC training splits mentioned in Section~\ref{sec5:dataset}. In any-to-any conversion, the same speaker encoder introduced in Section~\ref{sec3:a2a} is used to generate speaker vectors from mel-spectrograms. This setting is similar to the one in our prior work \cite{liu2018voice}, except that the i-vectors and learned speaker embedding vectors are used as speaker identity representations there. We use LibriTTS (train-clean-100 and train-clean-360) dataset together with the training set of VCTK corpus to train the conversion model.

The PPG extractor is obtained from a frame-wise phoneme recognizer, which is trained with the LibriSpeech dataset (960 hours). We first use the MFA to align the audio and transcripts at the phoneme level. Then the audio-text alignment information is used to obtain the frame-to-phoneme correspondence between mel-spectorgrams and phoneme sequences, with which it is possible to train a frame-wise phoneme recognizer. We regard the probability vectors after the last softmax layer as PPG features for one utterance.

\textbf{NonParaSeq2seq-VC}: This baseline approach is proposed by \cite{zhang2019non}, where disentangled content and speaker representations are extracted from acoustic features, and voice conversion is achieved by preserving the content representations of source utterances while replacing the speaker representations with the target ones. Since there is a speaker encoder which is jointly trained with the whole model, this approach can be trivially extended from any-to-many conversion to support any-to-any conversion.
We use the VCTK and CMU ARCTIC training splits to train the model for the any-to-many conversion setting and use the LibriTTS (train-clean-100 and train-clean-360) dataset together with the training set of VCTK corpus to train the model for the any-to-any conversion setting. We use the official implementation released by the authors\footnote{\url{https://github.com/jxzhanggg/nonparaSeq2seqVC\_code}} in this study.

\subsection{Objective evaluations}
\begin{figure}[t]
	\centering
	\subfloat[]{\includegraphics[width=0.4\textwidth]{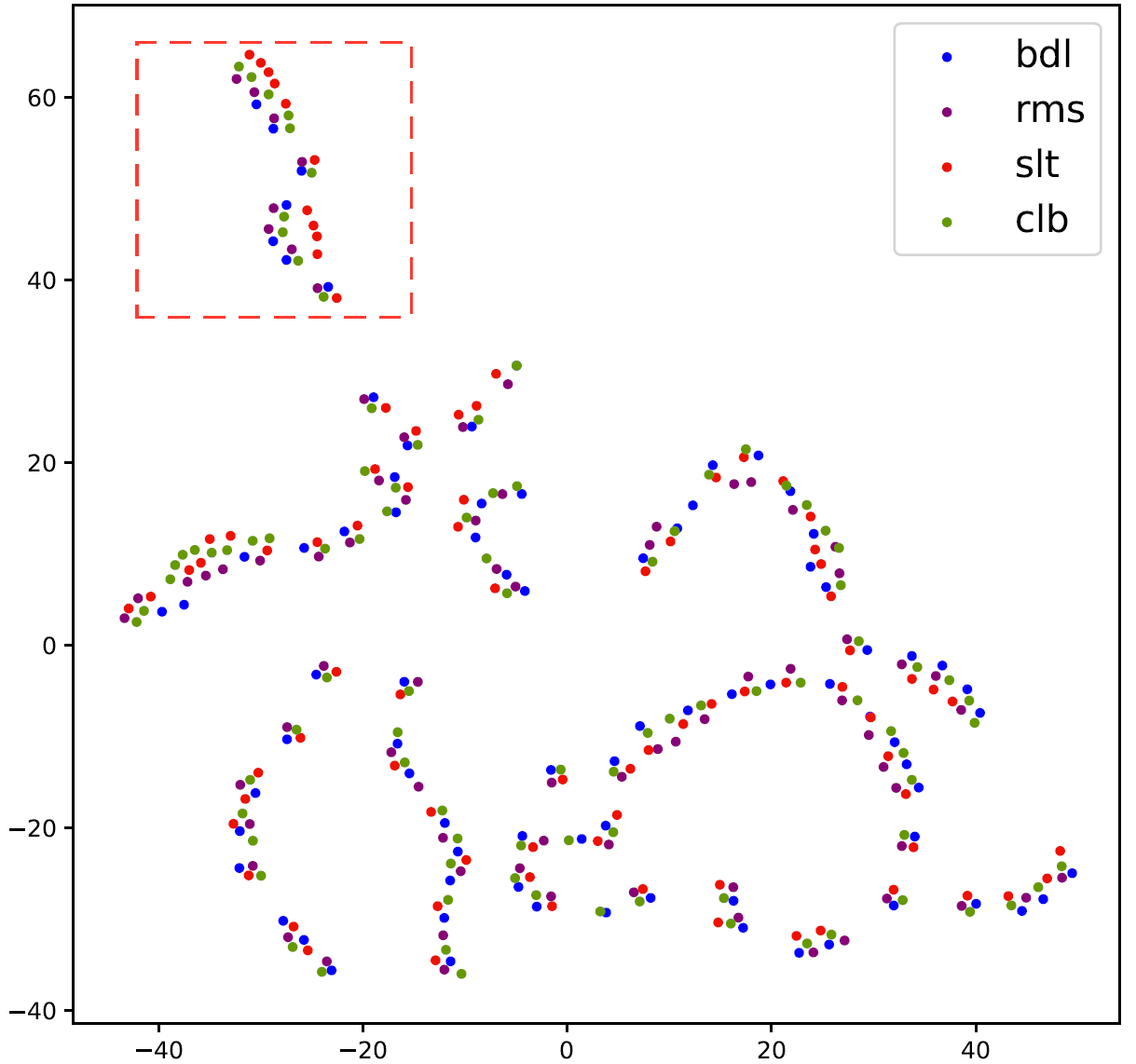}\label{chap5:tsnel}}
	\hfill
	\subfloat[]{\includegraphics[width=0.4\textwidth]{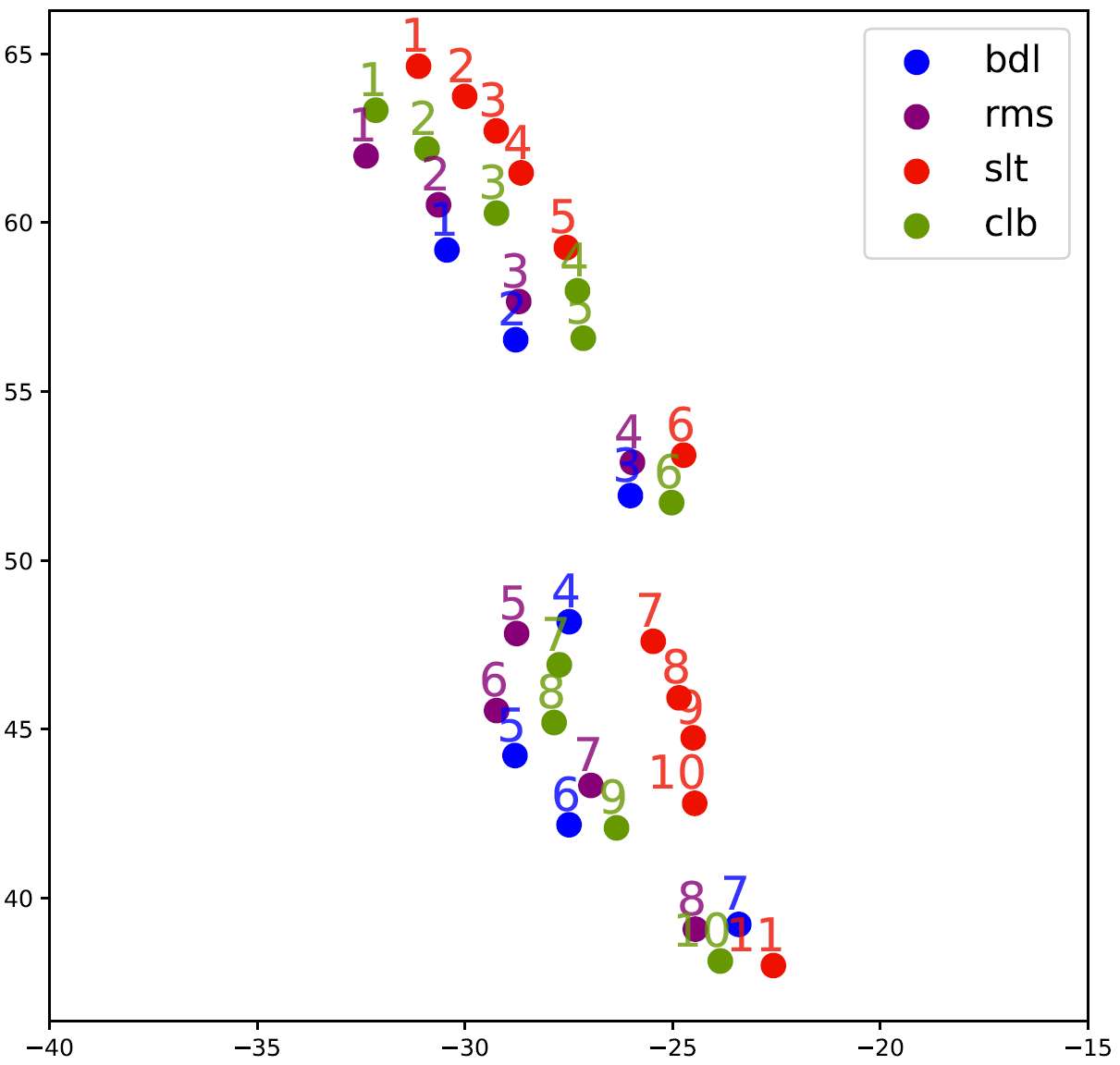}\label{chap5:tsnedetail}}
	\caption{(a) Visualization of bottle-neck features extracted by the BNE in the proposed BNE-Seq2seqMoL VC approach from the utterance ``arctic-a0001" in the CMU ARCTIC dataset with t-SNE. (b) Details of the dashed red box in (a), where the number represent sequential order of a frame in the bottle-neck features.}
	\label{chap5:tsne}
\end{figure} 
In our experiments, we find that the attention alignment matrices of the MoL attention are nearly diagonal, meaning that the proposed BNE-Seq2seqMoL system has little modeling capacity to effectively transform the duration and speaking rate.
Therefore, we only use mel-cepstrum distortion (MCD), root mean squared errors (F0-RMSE) and character/word error rate (CER/WER) from an ASR system as the metrics for objective evaluation. The MCD is used for evaluating spectral conversion, which is computed as:
\begin{align}
    \text{MCD[dB]} = \frac{10}{\text{log}10}\sqrt{2\sum_{d=1}^{K}(\text{MCC}_d^c - \text{MCC}_d^t)^2}
\end{align}
where MCC represents mel-cepstral coefficient, $K$ is the dimension of the MCCs, and $\text{MCC}_d^c$ and $\text{MCC}_d^t$ represent the $d$-th dimensional coefficient of the converted MCCs and the target MCCs, respectively. The Pyworld toolkit is used to extract MCCs in this paper, where we set $K=24$.

The F0-RMSE is used to evaluate the F0 conversion, which is computed as:
\begin{align}
    \text{F0-RMSE[Hz]} = \frac{1}{N}\sqrt{\sum_{i=1}^{N}(\text{F0}_i^c - \text{F0}_i^t)^2}
\end{align}
where $N$ is number of frames, $\text{F0}_i^c$ and $\text{F0}_i^t$ are F0 value at the $i$-th frame of the converted speech and target speech respectively.

We use a transformer-based end-to-end ASR engine\footnote{\url{https://github.com/espnet/espnet\_model\_zoo}} to compute CER and WER of the converted speech to evaluate its intelligibility. The ASR model is trained using the LibriSpeech (960 hours) dataset. The CER and WER for the CMU ARCTIC test set are 2.71\% and 4.30\%, respectively.

The objective evaluation results in the any-to-many voice conversion setting are shown in Table~\ref{obj_table1}. We can see that the proposed BNE-Seq2seqMoL approach achieves the best performance for all the four objective metrics on average.
The PPG-VC approach has the worst F0-RMSE among the four approaches, which verifies that conversion by frame-wise mapping has constrained capability to model prosody during the conversion. Comparing the NonParaSeq2seq-VC and the Seq2seqPR-DurIAN approaches, which have similar model architectures, we can see that the re-designed Seq2seqPR-DurIAN has superior results for all four metrics on average. The NonParaSeq2seq-VC has significantly worse CER and WER among the approaches, and preliminary listening test finds that there exists repeating, skipping and truncation phenomena in the converted speech. This implies that injecting a duration model, which provides explicit phone-level durational information, makes the conversion more robust. 
Comparing the results from PPG-VC with those from BNE-Seq2seqMoL shows that the auto-regressive property of the latter can boost VC performance across the four objective metrics. 

The objective evaluation results in the any-to-any voice conversion setting are presented in Table~\ref{obj_table2}. We can see that the PPG-VC approach has the best results in terms of MFC, CER and WER. The proposed BNE-Seq2seqMoL gave lowest F0-RMSE among the four approaches and obtains good results for MCD, CER and WER. 
Note that the models are trained using a combination of the VCTK training set and the large LibriTTS (train-clean-100, train-clean-360) dataset, which contains much more training data than in the any-to-many setting. This indicates that using more training data can overcome the deficiency of independent prediction across frames in the PPG-VC approach.

\begin{table*}[ht]
\centering
\caption{Ablation studies for the proposed BNE-Seq2seqMoL approach. ``IN" represents instance normalization and ``LSA" represents location-sensitive attention.}
\resizebox{\textwidth}{!}{
\begin{tabular}{c||c|c|c|c||c|c|c|c||c|c|c|c||c|c|c|c}
\thickhline
\multirow{2}{*}{\begin{tabular}[c]{@{}c@{}}Conversion \\ pair\end{tabular}} &
  \multicolumn{4}{c||}{BNE-Seq2seqMoL} &
  \multicolumn{4}{c||}{Without Log-F0\&UV} &
  \multicolumn{4}{c||}{Without IN} &
  \multicolumn{4}{c}{Use LSA} \\ \cline{2-17} 
 &
  \multicolumn{1}{c|}{MCD} &
  \multicolumn{1}{c|}{\begin{tabular}[c]{@{}c@{}}F0\\ RMSE\end{tabular}} &
  \multicolumn{1}{c|}{CER} &
  \multicolumn{1}{c||}{WER} &
  \multicolumn{1}{c|}{MCD} &
  \multicolumn{1}{c|}{\begin{tabular}[c]{@{}c@{}}F0\\ RMSE\end{tabular}} &
  \multicolumn{1}{c|}{CER} &
  \multicolumn{1}{c||}{WER} &
  \multicolumn{1}{c|}{MCD} &
  \multicolumn{1}{c|}{\begin{tabular}[c]{@{}c@{}}F0\\ RMSE\end{tabular}} &
  \multicolumn{1}{c|}{CER} &
  \multicolumn{1}{c||}{WER} &
  \multicolumn{1}{c|}{MCD} &
  \multicolumn{1}{c|}{\begin{tabular}[c]{@{}c@{}}F0\\ RMSE\end{tabular}} &
  \multicolumn{1}{c|}{CER} &
  \multicolumn{1}{c}{WER} \\ \hline
F-M & 6.87 & 44.55 & 3.00 & \textbf{4.29} & 6.91 & 49.92 & \textbf{2.23} & 5.14 & 6.85 & \textbf{41.65} & 3.94 & 6.46 & \textbf{6.81} & 46.64 & 4.31 & 7.19 \\ \hline
F-F & 6.99 & 45.44 & 5.12 & 6.97 & 7.05 & 47.24 & \textbf{3.97} & \textbf{6.12} & \textbf{6.93} & 45.77 & 4.90 & 6.96 & 7.00 & \textbf{44.03} & 4.28 & 7.20 \\ \hline
M-M & 6.94 & 50.37 & 4.72 & 7.22 & 7.00 & 54.20 & \textbf{3.93} & 6.80 & 7.04 & 52.81 & 5.52 & 8.51 & \textbf{6.90} & \textbf{49.71} & 4.53 & \textbf{6.44} \\ \hline
M-F & 7.18 & 56.31 & \textbf{3.61} & \textbf{5.76} & \textbf{7.13} & \textbf{47.80} & 5.09 & 7.25 & 7.23 & 60.56 & 5.45 & 7.75 & 7.15 & 58.65 & 4.89 & 6.84 \\ \hline
Average & 6.99 & \textbf{49.17} & 4.11 & \textbf{6.06} & 7.02 & 49.79 & \textbf{3.81} & 6.32 & 7.01 & 50.20 & 4.95 & 7.42 & \textbf{6.97} & 49.76 & 4.50 & 6.92 \\ \thickhline
\end{tabular}
}
\label{ablation}
\end{table*}

\subsection{Subjective evaluations}

Subjective evaluation in terms of both the naturalness and speaker similarity of converted speech are conducted\footnote{Audio demo and source codes can be found in \url{https://liusongxiang.github.io/BNE-Seq2SeqMoL-VC/}.}. 
We use the 5 point Likert scale for testing with mean opinion score (MOS) (1-bad, 2-poor, 3-fair, 4-good, 5-excellent) regarding both naturalness and speaker similarity evaluations. In the MOS tests for evaluating naturalness, each group of stimuli contains recording samples from the target speakers, which are randomly shuffled with the samples generated by the four comparative approaches, before they are presented to listeners. In the MOS similarity tests, converted speech samples are directly compared with the recording samples of the target speakers. 10 utterances from the CMU ARCTIC test set are presented for each conversion pair (i.e., F-M, F-F, M-M and M-F). We invited 35 raters who are proficient in English to participate in the evaluations in a quiet room and they were asked to use headphones during the tests. The raters were allowed to replay each sample as many times as necessary and change their ratings of any sample before submitting their results.

The subjective MOS evaluation results for the any-to-many voice conversion setting are shown in Table~\ref{sub_table1}.
We can see that the proposed BNE-Seq2seqMoL approach achieves the best average results in terms of both naturalness and speaker similarity. More specifically, for the pairs of F-F, M-M, F-M, the proposed model achieve better performance. While for the pair of M-F, Seq2seqPR-DurIAN provides better result. The possible reason might be that the simple linear transformation in logarithm scale used by the BNE-Seq2seqMoL approach (see Eq. (13)) can not adequately model the actual male-to-female pitch conversion; and this can also be reflected by the high F0 RMSE (i.e., 7.18) between the converted and reference speech as shown in Table I.
In any-to-any conversion, we only conduct MOS tests for the proposed BNE-Seq2seqMoL approach.
The results are presented in Table~\ref{sub_table2}. According to the absolute MOS values, we can see that the proposed approach also achieves good VC performance in the one-shot/few-shot voice conversion setting.

\subsection{Cross-speaker property of bottle-neck features}
To explore the property of the bottle-neck features extracted by the BNE in the BNE-Seq2seqMoL approach, t-distributed Stochastic Neighbor Embedding (t-SNE)~\cite{maaten2008visualizing} is used to visualize bottle-neck features. t-SNE is a non-linear dimensionality reduction technique that is widely used to embed high-dimensional data into a space of two or three dimensions. The goal is to find a faithful representation of those high-dimensional data in a low-dimensional space. 

Fig.~\ref{chap5:tsnel} is the 2-dimensional t-SNE visualization of the bottle-neck features of the utterance ``arctic-a0001" of the four speakers (bdl, rms, slt and clb). The 256-dimensional bottle-neck features are fed into t-SNE and then the result is obtained after five thousand iterations. In the figure, each bottle-neck feature frame is represented by a dot. The four colors represent the four speakers. We can see a strong degree of clustering effect of the bottle-neck features across different speakers. The details of the red dashed box in Fig.~\ref{chap5:tsnel} are depicted in Fig.~\ref{chap5:tsnedetail}, where the numerical indices represent sequential frame order of the first several frames in the bottle-neck features. A similar manifold pattern can be observed across the four examined speakers, which shows the goodness of cross-speaker property of the bottle-neck features used in the proposed BNE-Seq2seqMoL approach.
This can also be a reasonable explanation of the superior VC performance of the BNE-Seq2seqMoL approach since obtaining speaker-agnostic content representations from the source speech makes the synthesis module focus more on generating the target voice.

\subsection{Ablation Studies}
In this section, ablation studies are conducted to validate the effectiveness of the feature selection and model design strategies in the proposed BNE-Seq2seqMoL approach. Specifically, three ablation studies are conducted: 1) dropping the Log-F0 and UV features and only use bottle-neck features as input to the synthesis module; 2) dropping the instance normalization layers in the pitch encoder; 3) using the location-sensitive attention (LSA) instead of the location-relative MoL attention.

The objective evaluation of the ablation studies are shown in Table~\ref{ablation}. We observe that the proposed feature selection and model design for the BNE-Seq2seqMoL approach obtains the best F0-RMSE and WER results, and achieve near-the-best MCD and CER results. This validates the effectiveness of the feature selection and model design in the BNE-Seq2seqMoL approach.

\subsection{Training and Inference Speed}

In all experiments, training processes are stopped early if the validation losses do not decrease for five epochs. We use Pytorch toolkit to implement all models without any hard-ware optimization. The computation platform information is: NVIDIA Tesla M40 GPU and Intel(R) Xeon(R) CPU E5-2680(v4) @ 2.40GHz. We measure training time of the Seq2seq-DurIAN and the BNE-Seq2seqMoL approaches in the any-to-many setting. Training 10 epochs of the phoneme recognizer takes about 40 hours using 8 GPUs. The synthesis module in the Seq2seq-DurIAN system is trained 200 epochs on 1 GPU, taking 44 hours. The synthesis module in the BNE-Seq2seqMoL apporoach is trained 56 epochs on 1 GPU, taking 28 hours.

The inference speed on GPUs for all the compared four systems for any-to-many VC is also measured using 50 samples from the testing set. Since we use an open-sourced auto-regressive WaveRNN model as the vocoder to generate waveform, we exclude the WaveRNN inference time to prevent it  from dominating the computation. Inference real-time factors (RTF) are presented in Table VI. The beam size is 10 for the Seq2seq-DurIAN and NonParaSeq2seq-VC systems.

The bottle-neck feature extractor in the BNE-Seq2seqMoL approach has several BiLSTM layers, which compute input frames recurrently. To explore the possibility of accelerating the inference speed for the BNE-Seq2seqMoL approach, we partition the input mel spectrograms evenly into $N \in [{2, 4, 6, 8, 16}]$ segments along the temporal axis and compute the bottle-neck features in a batch mode. The bottle-neck features of the $N$ segments are then concatenated temporally before being fed into the remaining parts of the system. The RTF results are presented in Table VII, where ``Folding x$N$" refers to partitioning into $N$ segments. We can see that partitioning a mel spectrogram into 2 segments can get the greatest inference speed gain; and our preliminary listening test shows that this partitioning operation (i.e., ``folding x2") does not hurt the conversion performance.

\begin{table}[!h]
	\centering
	\caption{Inference speed information. Pytorch implementation without hardware optimization for an Nvidia Tesla M40 GPU and Intel(R) Xeon(R) E5-2680(v4) CPU @ 2.40GHz.}
	\label{tableMCD}
	\begin{tabular}{cc}
		\thickhline
		\textbf{System} & RTF (GPU) \\ \hline
		PPG-VC & 0.039 \\
		NonParaSeq2seq-VC & 0.117 \\
		Seq2seqPR-DurIAN & 3.329 \\
		BNE-Seq2seqMoL & 0.245 \\ 
		\thickhline
	\end{tabular}
\end{table}

\begin{table}[!h]
	\centering
	\caption{Inference speed information of the BNE-Seq2seqMoL approach using different folding rate. We use a Pytorch implementation without hardware optimization for an Nvidia Tesla M40 GPU and Intel(R) Xeon(R) E5-2680(v4) CPU @ 2.40GHz.}
	\label{tableMCD}
	\begin{tabular}{cc}
		\thickhline
		\textbf{Different folding rate} & RTF (GPU) \\ \hline
		Without folding & 0.245 \\
		Folding x2 & 0.221 \\
		Folding x4 & 0.224 \\
		Folding x8 & 0.228 \\
		Folding x16 & 0.242 \\
		\thickhline
	\end{tabular}
\end{table}

\section{Conclusion}
\label{sec6}
In this paper, we re-design a prior approach \cite{liu2020transferring} to achieve a robust non-parallel seq2seq any-to-many VC approach. The novel approach concatenates a seq2seq phoneme recognizer (Seq2seqPR) and a multi-speaker duration informed attention network (DurIAN) for synthesis. Extension is also made on this approach to enable support of any-to-any voice conversion. Thorough examinations including objective and subjective evaluations are conducted for this model in any-to-many, as well as any-to-any settings.

To overcome the deficiencies of the PPG-based and non-parallel seq2seq any-to-many VC approaches, we further proposed a new any-to-many VC approach, which combines a bottle-neck feature extractor (BNE) with an MoL attention-based seq2seq synthesis model. This approach can easily be extended to any-to-any VC. Objective and subjective evaluation results show its superior VC performance in both any-to-many and any-to-any VC settings. Ablation studies have been conducted to confirm the effectiveness of feature selection and model design strategies in the proposed approach. 
The proposed BNE-Seq2seqMoL approach has successfully shortened the sequence-to-sequence VC pipeline to contain only an ASR encoder and a synthesis decoder. However, it still uses spectral features (i.e., mel spectrograms) as intermediate representations and relies on an independently trained neural vocoder to generate the waveform. This may reduce the synthesis quality, which be avoided by jointly training the whole VC pipeline in an end-to-end manner (i.e., waveform-to-waveform training). In the future, we will also explore the proposed approach in terms of source style transfer and emotion conversion.

\bibliographystyle{IEEEtran}
\bibliography{./bib/is20,./bib/local,./bib/icassp20,./bib/is19}

\ifCLASSOPTIONcaptionsoff
  \newpage
\fi

\begin{IEEEbiographynophoto}{Songxiang Liu}
received his B.S. degree in Automation from the College of Control Science and Engineering, Zhejiang University (ZJU), Hangzhou, China, in 2016. He is currently a Ph.D candidate at the Human Computer Communications Lab (HCCL) in the Chinese University of Hong Kong, Hong Kong SAR, China. His research focuses on voice conversion, accent conversion, audio adversarial attack \& defense, text-to-speech synthesis and automatic speech recognition.
\end{IEEEbiographynophoto}

\begin{IEEEbiographynophoto}{Yuewen Cao}
received her B.S. degree in communication engineering from Huazhong University of Science \& Technology, Wuhan, China, in 2017. She is currently pursuing her Ph.D degree at the Human Computer Communications Lab (HCCL) in the Chinese University of Hong Kong, Hong Kong SAR, China. Her research interests include speech synthesis and voice conversion.
\end{IEEEbiographynophoto}

\begin{IEEEbiographynophoto}{Disong Wang}
received his B.S. in Mathematics \& Physics Basic Science from University of Electronic Science and Technology of China (UESTC) in 2015, and M.E. in Computer Applied Technology from Peking University (PKU) in 2018. He is currently a Ph. D. candidate at the Human Computer Communications Lab (HCCL) in the Chinese University of Hong Kong (CUHK). His research interests include voice conversion, text-to-speech synthesis, automatic speech recognition and their applications to non-standard speech, such as accented voice and dysarthric speech.
\end{IEEEbiographynophoto}

\begin{IEEEbiographynophoto}{Xixin Wu}
received the B.S., M.S. and Ph.D. degrees from Beihang University, Tsinghua University and The Chinese University of Hong Kong, China, respectively. He has been a Research Assistant at the Machine Intelligence
Laboratory, Cambridge University Engineering Department. His research interests include speech synthesis, voice conversion speech recognition and neural network uncertainty.
\end{IEEEbiographynophoto}

\begin{IEEEbiographynophoto}{Xunying Liu}
received the bachelor’s degree from Shanghai Jiao Tong University, the Ph.D. degree in speech recognition, and the M.Phil. degree in computer speech and language processing both from the University of Cambridge, Cambridge, UK. He has been a Senior Research Associate at the Machine Intelligence Laboratory, Cambridge University Engineering Department, and from 2016 an Associate Professor in the Department of Systems Engineering and Engineering Management, the Chinese Univer sity of Hong Kong. He received the Best Paper Award at ISCA Interspeech 2010. His current research interests include large vocabulary continuous speech
recognition, language modelling, noise robust speech recognition, speech synthesis, speech and language processing. He is a Member of ISCA.
\end{IEEEbiographynophoto}

\begin{IEEEbiographynophoto}{Helen Meng}
(F’13) received the S.B., S.M., and Ph.D. degrees in electrical engineering from the Massachusetts Institute of Technology, Cambridge, MA, USA. She joined The Chinese University of Hong Kong in 1998, where she is currently a Professor with the Department of Systems Engineering and Engineering Management. She was also the Associate Dean of Research of the Faculty of Engineering between 2005 and 2010. Her research interests include human–computer interaction via multimodal and multilingual spoken language systems, speech retrieval technologies, and computer-aided pronunciation training. She served as the Editor-in-Chief of the IEEE TRANSACTIONS ON AUDIO, SPEECH, AND LANGUAGE PROCESSING from 2009 to 2011. She was an Elected Board Member of the International Speech Communication Association as well as an Elected Member of the IEEE Signal Processing Society Board of Governors. She is a Fellow of the HKCS, HKIE, and ISCA.
\end{IEEEbiographynophoto}







\end{document}